\documentclass[12pt]{article}
\usepackage{epsfig}
\begin{document}

\author{S.~Bucikiewicz, L.~D\c{e}bski, W. Florek}
\title{Application of algebraic combinatorics to finite spin systems
 with dihedral symmetry }
\date{Institute of Physics, A.~Mickiewicz University,\\ 
ul. Umultowska 85, 61--614 Pozna\'n, Poland}
\maketitle
\begin{abstract}
\baselineskip=1.3\baselineskip
  Properties of a given symmetry group $G$ are very important in
investigation of a physical system invariant under its action. In the case
of finite spin systems (magnetic rings, some planar macromolecules) the
symmetry group is isomorphic with the dihedral group $\mathrm{D}_N$.
In this paper group-theoretical `parameters' of such groups are determined,
especially decompositions of transitive representations into irreducible
ones and double cosets. These results are necessary to construct matrix
elements of any operator commuting with $G$ in an efficient way. The 
approach proposed can be usefull in many branches of physics, but 
here it is applied to finite spin systems, which serve as models for
mesoscopic magnets.
\end{abstract}

\vspace*{5mm}

\noindent Corresponding author:\\
\hspace*{1cm}  Wojciech Florek\\
\hspace*{1cm}  Uniwersytet im. A. Mickiewicza\\
\hspace*{1cm}  Instytut Fizyki\\
\hspace*{1cm}  Zak{\l}ad Fizyki Komputerowej\\
\hspace*{1cm}  ul. Umultowska 85\\
\hspace*{1cm}  61--614 Pozna\'n\\
\hspace*{1cm}  e-mail: Wojciech.Florek@spin.amu.edu.pl\\
\hspace*{1cm}  tel.: (0--61) 827 30 33\\
\hspace*{1cm}  fax: (0--61) 825 77 58

\clearpage
\baselineskip=1.3\baselineskip

\section{Introduction}
One of standard ways of quantum magnets modeling is to consider a
finite set of $N$ nodes carrying spins ${\mathbf s}_j$, $1\leq j\le N$,
with a spin number $s$. At present such models are also used to 
investigate properties of magnetic macromolecules as 
Mn$_{12}$acetate \cite{andrea}. The so-called {\em Ising states}\/ or
{\em configurations}\/, {\it i.e.}\ tensor products 
 \begin{equation} \label{state}
  \mu=|m_1,m_2,\dots,m_N\rangle \,,\qquad -s\le m_j\le s\,,
 \end{equation}
of one-spin states $|m_j\rangle$ form a natural basis $\mathcal{B}$ of 
the quantum space of states $L$ with the dimension $\dim L=(2s+1)^N$. 
In a general case, however, these states are {\em not\/} eigenvectors 
of the Heisenberg-like Hamiltonians. Therefore, it is necessary to solve 
an eigenproblem with $\dim L$ equations what, even using contemporary 
computers, is almost impossible for $N\approx 30$ and $s>1$. This is 
especially important when one wants to investigate spin correlations in 
the ground state and some of the first excited states of antiferromagnetic 
systems, since many methods allow to determine eigenvalues, but it is 
much more difficult to find eigenvectors. Among others, it is caused by 
the numerical instability of necessary algorithms in the case of a very
large number of linear equations \cite{nrc}. The problem is partially 
solved by taking into account operators commuting with a given 
Hamiltonian: {\it e.g.}\ the $z$-th component of the total spin $S^z$ and 
irreducible representations of the Hamiltonian symmetry group 
\cite{BF,hr12}. If the spin interactions are isotropic, then also the 
total spin number may be taken into account. However one has to find 
solutions of a set of homogeneous linear equations arising from the 
eigenproblem for ${\mathbf S}^2$ (in finite spin system eigenvalues of
${\mathbf S}^2$ are known). In many interesting cases, for example 
investigating the ground state of bipartite antiferromagnets, these 
solutions can be found {\it exactly}\/ if one is able to cope with very 
large integers. Such tools are provided by multiple (or arbitrary) 
precision libraries like the GNU Multiple Precision Arithmetic Library 
({\sf GMP} or {\sf GNU MP} in short) written and developed by 
T.~Granlund \cite{gmp}. However, such procedures work efficiently if the 
matrix of an operator (the squared total spin ${\mathbf S}^2$ or the 
Hamiltonian ${\mathcal H}$ in the presented considerations) are written 
in the symmetry adapted basis. 

The considerations presented are illustrated by examples related with
finite spin models, but a similar method can be applied in any case when:
 \begin{itemize}
  \item the space of states $L$ is spanned over orthonormal vectors
    (\ref{state}), {\it i.e.}\ it is the $N$th tensor power of 
    the $n$-dimensional one-particle space;
  \item the considered operator ${\mathcal H}$, acting in $L$, commutes 
    with a given permutation representation $P$ of a group $G$ (the 
    Hamiltonian symmetry group).
 \end{itemize}

Let a given group $G$ permutes vectors of a (finite) basis $\mathcal{B}$ 
of the quantum state space $L$. In the previous paper \cite{prev} general 
formulas for vectors of the irreducible basis and matrix elements of any 
operator commuting with all permutation operators $P(g)$, $g\in G$, have 
been presented. The aim of the present work is to investigate the dihedral 
groups $\mathrm{D}_N$ and obtain general analytical expressions in this 
case. These groups have been chosen, since they describe symmetry of 
molecular rings with $N$ identical spins $s$ \cite{andrea}. In fact, these 
rings are also invariant under the reflection $\sigma_h$ in the horizontal 
plane, so the groups $\mathrm{D}_{Nh}$ should be considered. These groups 
are important, for example, constructing symmetry coordinates of vibrating 
molecules or crystals, but in the case of spin models the action of 
$\sigma_h$ is trivial. Considering planar molecules one should introduce 
groups $\mathrm{C}_{Nv}$, which are elementary `bricks' for finite space 
groups \cite{jpa}. However, groups $\mathrm{D}_N$ and $\mathrm{C}_{Nv}$ 
are isomorphic and the abstract group $\mathrm{D}_N$ is considered as a 
representative of isomorphic groups with the following generation 
relations \cite{cox}:
 \begin{equation}\label{dngrel}
   \mathrm{D}_N=\langle C_N,U_0\rangle\,,\qquad
     C_N^N=U_0^2=(C_N^jU_0)^2=E\,,\;0\le j<N\,.
  \end{equation}
 \begin{figure}
  \begin{center}\epsfxsize 7cm \epsfbox{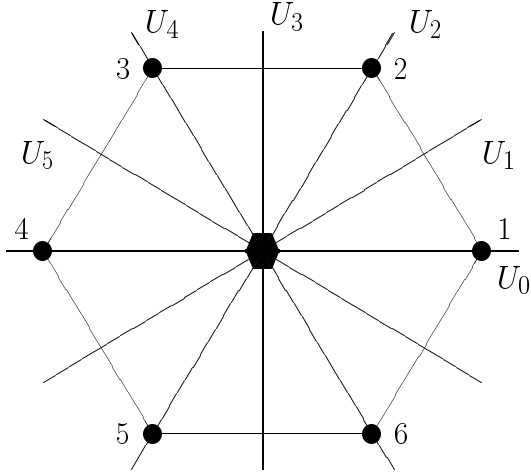} \end{center}
  \caption{Numbering of ring nodes and elements of the group 
    $\mathrm{D}_6$ (the six-fold axis $C_6$ is denoted by the full 
    hexagon)}\label{group}
 \end{figure}
 The dihedral group $\mathrm{D}_N$ consists of $2N$ elements: $N$ powers 
of the $N$-fold rotation $C_N$ and perpendicular to it the twofold rotations 
$U_j=C_N^jU_0$, $0\le j<N$. It is easy to write down general formulas for 
multiplication (all calculations are done $\bmod\, N$):
 \begin{equation}\label{mulrul}
   C_N^jC_N^k=C_N^{j+k}\,,\; C_N^jU_k=U_{j+k}\,,\;
   U_jC_N^k  =U_{j-k}\,,\; U_jU_k    =C_N^{j-k}\,.
 \end{equation}
 The labeling scheme for $N=6$ is presented in Fig.~\ref{group}. 

The dihedral groups are considered in Sec.~\ref{dng}; there are many 
differences between cases of $N$ odd and even, so the results are presented
separately. In many actual applications one deals with bipartite
antiferromagnets \cite{andrea}, so the case of $N$ even seems to be
more important, however the other case ($N$ odd) is discussed for
three, at least, reasons: (i) the ground state energy of infinite linear
antiferromagnets is approximated from below and above by those obtained
in the cases of even- and odd-sized rings, respectively \cite{hr12};
(ii) there are still some open problem related with linear ferromagnets
({\it e.g.} critical exponents near $T=0$\,K \cite{lechu}); in this case
the parity of $N$ is (almost) irrelevant; (iii) the presented 
considerations and results are more complete. In Sec.~\ref{example}
a partial solution for six spins $s=1$ (a ring invariant under 
$\mathrm{D}_6$) is discussed. The paper ends with some conclusions.

\section{Dihedral groups $\mathrm{D}_N$}\label{dng}

Let us recall the formulas obtained in the previous paper \cite{prev} to
check what group properties should be determined. The irreducible basis 
consists of vectors $|U\,\mu;\Gamma r \gamma\rangle$, where:
 \begin{itemize}
   \item  $U\subseteq G$ is a subgroup of $G$ representing a conjugacy 
     class $\widetilde{U}$; 
   \item $\mu$ is a representative of an orbit of type $U$; 
   \item $\Gamma$ is an irreducible representation (an irrep)
     of $G$, the repetition index $r$ distinguishes copies of $\Gamma$ 
     and $\gamma=1,2,\dots,[\Gamma]$ labels vectors of $\Gamma$.
  \end{itemize} 
 The orbit $G(\mu)$ can be identified more precisely if one introduces
a label denoting a non-ordered partition $[k]$ such that 
$\mu\in\mathcal{B}_{[k]}$. Note that $[k]$ identifies also the 
magnetization $M$. However, the formula for vectors of the irreducible 
basis does not depend on $[k]$ and is given as \cite{prev}
 \begin{equation}\label{irrbas}
   |U\,\mu;\Gamma r \gamma\rangle = \sum_{j=1}^{|G/U|} 
   b^{\Gamma r \gamma}_j(U)|U\,\mu\,g_j\rangle\,,
 \end{equation}
 where the elements $g_j$, $1\le j\le|G/U|$ with $g_1=E$, represents the 
left cosets $gU\in G/U$. The coefficients $b^{\Gamma r\gamma}_j$, 
determining a unitary transformation, can be expressed by matrix elements 
of the irrep $\Gamma$ and the reduction coefficients 
$A^{U\Gamma}_{\gamma r}$: 
 \begin{equation}\label{cofbe}
   b_j^{\Gamma r \gamma}(U)=
    \left({{[\Gamma]|U|}\over{|G|}}\right)^{\!1/2}
    \sum_{\beta=1}^{[\Gamma]} \Gamma^*_{\gamma\beta}(g_j)
      (A^{U\Gamma}_{\beta r})^*\,.
 \end{equation}
 Matrix elements of any operator, {\it e.g.} a Hamiltonian ${\mathcal H}$,
commuting with all $P(g)$ are grouped in blocks labeled by $\Gamma$ and 
depend only on $U$, $\mu$ and $r$: 
  \begin{equation} \label{mael}
    h_{U \mu r,V \nu s}(\Gamma)=\left({|U|\over |V|}\right)^{\!1/2}\,
    \sum_{j=1}^{|G/U|} \langle U\,\mu\,g_j|{\mathcal H}|V\,\nu\,E\rangle
        B^{UV\Gamma}_{rs}(g_j)\,,
 \end{equation}
  where 
 \begin{equation} \label{bcof}
    B^{UV\Gamma}_{rs}(g_j)=
     \sum_{\alpha,\beta} (A^{V\Gamma}_{\alpha s})^* 
       \Gamma_{\alpha\beta}(g_j) A^{U\Gamma}_{\beta r} \,.
 \end{equation}
 It is helpful to consider non-ordered partitions $[k]$, because matrix 
elements are equal to zero if partitions $[k]$ and $[k']$ are not joined 
by an edge in the graph introduced in Sec.~5 of the previous paper 
\cite{prev}. Introducing the notion of double cosets $VgU$ this formula 
can be modified and written as: 
 \begin{equation}\label{Hgd} 
   h_{U \mu r,V \nu s}(\Gamma)=\left(|U||V|\right)^{-1/2}\:
   \sum_{d=1}^{|V\backslash G/U|} |Vg_dU| B^{UV\Gamma}_{rs}(g_d) 
      \langle U\,\mu \,g_d|{\mathcal H}|V\,\nu\,E\rangle\,,
 \end{equation}
 where elements $g_d$ represent double cosets $Vg_dU$, or orbits of
the natural action $_V(G/U)$.  

All structures appearing in these formulas (cosets, irreps, reduction
coefficients {\it etc.}) in the case $G=\mathrm{D}_N$ are discussed in the 
following subsections.

\subsection{Classes and irreps}

 The multiplication rules (\ref{mulrul}) in the dihedral group 
$\mathrm{D}_N$ lead immediately to classes of conjugated elements:
 \begin{eqnarray*}
   N\;\mathrm{odd}:&&\{E\},\; 
     \{C_N^j,C_N^{N-j}\},\quad j=1,2,\dots,\frac{N-1}{2}\,,\quad
     \{U_0,U_1,\dots,U_{N-1}\}\,;\\  
   N\;\mathrm{even}:&&\{E\},\; 
     \{C_N^j,C_N^{N-j}\},\quad j=1,2,\dots,\frac{N}{2}-1\,,\\
       && \{C_2=C_N^{N/2}\},\quad
     \{U_0,U_2,\dots,U_{N-2}\}\,\quad\{U_1,U_3,\dots,U_{N-1}\}.
  \end{eqnarray*}
 Therefore, there are $(N+3)/2$ classes (and irreps) for $N$ odd and 
$N/2+3$ for $N$ even. In the special cases, $N=1,2$, the groups $D_N$ are 
Abelian. 

In both cases, $N$ odd and even, there are two one-dimensional irreps, 
denoted $A_1$ (the unit irrep) and $A_2$, with characters
 \begin{eqnarray}
   \chi^{A_1}(g)=1\,, && \forall\;g\in\mathrm{D}_N; \label{a1rep}\\
   \chi^{A_2}(C_N^j)=1\,,\;\chi^{A_2}(U_j)=-1\,, && j=0,1,\dots,N-1\,.
     \label{a2rep}
 \end{eqnarray}
 For $N$ even there are two more one-dimensional irreps, denoted $B_1$ 
and $B_2$, with characters ($j=0,1,\dots,N-1$)
 \begin{eqnarray}
   \chi^{B_1}(C_N^j)=(-1)^j\,,&&\;\chi^{B_1}(U_j)=(-1)^j \,; 
        \label{b1rep}\\
   \chi^{B_2}(C_N^j)=(-1)^j\,,&&\;\chi^{B_2}(U_j)=-(-1)^j \,.
        \label{b2rep}
  \end{eqnarray}
 Note that the element $C_2=C_N^{N/2}$ behaves in different ways for 
$N=4M$ and $N=4M+2$. In the first case one obtains
$\chi^{B_1}(C_2)=\chi^{B_2}(C_2)=1$, whereas in the second case this 
character equals $-1$.

 The other irreps, denoted $E_l$, $l=1,2,\dots,N_E$, are two-dimensional; 
there are $N_E=(N-1)/2$ and $N_E=N/2-1$ such irreps for $N$ odd and even, 
respectively. Their characters are given by the same formulas
 \begin{equation}
    \chi^{E_l}(C_N^j)=2\cos(2\pi jl/N)\,,\quad
      \chi^{E_l}(U^j)=0\,.\label{elrep}
 \end{equation}
 To complete the presentation, matrices of the representations $E_l$ in 
the standard basis $\{|Ex\rangle, |Ey\rangle\}$ for the generators $C_N$ 
and $U_0$ are written down:
 \begin{equation}\label{repemat}
   E_l(C_N)=\left(\begin{array}{rr}
         \cos(2\pi jl/N) & -\sin(2\pi jl/N) \\
         \sin(2\pi jl/N) & \cos(2\pi jl/N)
       \end{array}\right)\,,\quad
   E_l(U_0)=\left(\begin{array}{rr}
            1 & 0 \\ 0 & -1
       \end{array}\right)\,.
 \end{equation}
 
\subsection{Subgroups and conjugated subgroups}\label{subgr}
 Let $N$ be written as a product of $r$ different primes $p_i$:
  $$
       N=\prod_{i=1}^r p_i^{\alpha_i}\,,\quad \alpha_i>0\,.
  $$
 Then there are $\mathcal{D}=\prod_{i=1}^r (\alpha_i+1)$ different 
divisors $k$ of $N$, each of them can be written as ($k=1$ corresponds to 
all $\beta_i=0$)  
  $$
    k=\prod_{i=1}^r p_i^{\beta_i}\,,\quad 0\le\beta_i\le\alpha_i\,.
  $$

The first family of subgroups consists of the cyclic groups $\mathrm{C}_k$,
where $k$ is a divisor of $N$. These groups are generated by elements
$C_N^\kappa$, where $\kappa=N/k$ is the so-called co-divisor of 
$k$. For $k=1$ one obtains the group $\mathrm{C}_1=\{E\}$, which has
no generators; in this case $\kappa=N$. Each subgroup $\mathrm{C}_k$ 
contains $k$ elements in the form $C_N^{p\kappa}\equiv C_k^p$, 
$p=0,1,\dots,k-1$. Since these groups are normal divisors of 
$\mathrm{D}_N$, then each $\mathrm{C}_k$ is conjugated to itself only. 

\begin{figure}
  \begin{center}\epsfxsize 5cm \epsfbox{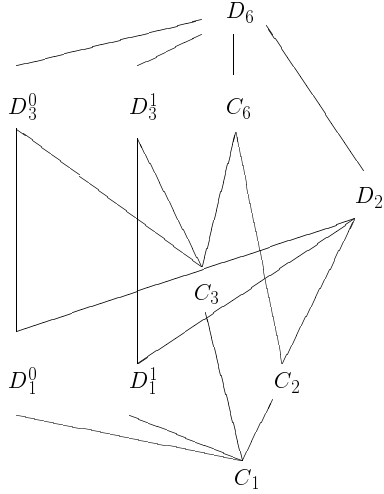} \end{center}
  \caption{Lattice of conjugated subgroups for the dihedral group 
   $\mathrm{D}_6$; lines correspond to the group-subgroup relations}
  \label{ld6}
\end{figure}

The second family of subgroups is formed by the dihedral groups 
$\mathrm{D}_k$.  However, for each $k$, {\it i.e.} for each generator 
$C_N^\kappa$, one has freedom in a choice of two-fold generator $U_j$. 
The subgroups isomorphic to $\mathrm{D}_k$ are generated by $C_N^\kappa$ 
and one of the rotations $U_0,U_1,\dots,U_{\kappa-1}$. These subgroups 
will be denoted hereafter as 
 $$
   \mathrm{D}_k^q=\langle C_N^\kappa,U_q\rangle\,, 
      \quad q=0,1,\dots,\kappa-1\,.
 $$
 Each of these groups contains, except for the rotations $C_N^{p\kappa}$, 
twofold axes $U_{q+p\kappa}$, $p=0,1,\dots,k-1$. In the case of odd 
$\kappa$ these elements belong to the same conjugacy class, so all 
subgroups $\mathrm{D}_k^q$ are conjugated. This conjugacy class is 
represented by $\mathrm{D}_k=\langle C_N^\kappa,U_0\rangle$. When $\kappa$ 
is even, then all two-fold $U_j$ rotations in the subgroup 
$\mathrm{D}_k^q$ have indices with the same parity. Therefore, there are 
two classes of conjugated subgroups: the first is represented by 
$\mathrm{D}_k^0=\langle C_N^\kappa,U_0\rangle$ and the second by 
$\mathrm{D}_k^1=\langle C_N^\kappa,U_1\rangle$. If $N$ is even then it can 
be written as $2^\alpha N'$, where $N'$ is odd. In the same way all 
divisors $k$ of $N$ can be written as $2^\beta k'$. If $\beta=\alpha$ then 
corresponding $\kappa=N'/k'$ is odd and, therefore, there is only one class
of conjugated subgroups $\mathrm{D}_k$. Two separated classes,
$\mathrm{D}_k^0$ and $\mathrm{D}_k^1$, are obtained in the other cases,
{\it i.e.}\ when $\beta<\alpha$ and $\kappa=2^{\alpha-\beta}N'/k'$. 
Let $\mathcal{D}'=\mathcal{D}/(\alpha+1)$ be the number of divisors of 
$N'$ ($\mathcal{D}'=\mathcal{D}$ if $N$ is odd) then:
  \begin{itemize}
    \item there are $\mathcal{D'}$ classes of dihedral subgroups in 
       type $\mathrm{D}_k$;
    \item in each type $\mathrm{D}_k^0$ and $\mathrm{D}_k^1$ there are 
     $\alpha\mathcal{D}'$ classes of dihedral subgroups. 
  \end{itemize}
 Therefore, there are $(\alpha +1)\mathcal{D}'$ cyclic subgroups and 
$(2\alpha +1)\mathcal{D}'$ conjugacy classes of dihedral subgroups. 
The lattice of conjugated subgroups for the group $\mathrm{D}_6$ is 
presented in Fig.~\ref{ld6}. 

 The theorem presented in Sec.~2.3 of the previous paper \cite{prev} 
immediately excludes some classes of subgroups if vectors of the considered 
basis are mappings, as in the case of magnetic configurations. The first 
excluded subgroup is the cyclic group $\mathrm{C}_N$, since there is only 
one orbit $X=\{0,1,\dots,N-1\}$, so $\prod_j \Sigma_{X_j}=\Sigma_{X}$ and 
$P(\mathrm{D}_N)\cap \Sigma_{X}=P(\mathrm{D}_N)\ne P(\mathrm{C}_N)$. 
In the case of $N$ even two other subgroups are excluded: the dihedral 
subgroup $\mathrm{D}_{N/2}^1$ and the cyclic subgroup $\mathrm{C}_{N/2}$. 
In the first case only one orbit $X$ is again obtained, whereas 
$\mathrm{C}_{N/2}$ decomposes $X$ into subsets $X_0$ and $X_1$ 
containing nodes with even and odd indices, respectively. Then one obtains 
$P(\mathrm{D}_N)\cap(\Sigma_{X_0}\times \Sigma_{X_1})=P(\mathrm{D}_{N/2}^0)
\ne P(\mathrm{C}_{N/2})$. Therefore, these subgroups can be omitted in the 
further considerations of spin configurations. For example, when $N=6$ 
subgroups $\mathrm{C}_6$, $\mathrm{D}_3^1$, and $\mathrm{C}_3$ are excluded.

\subsection{Cosets}\label{cosets}
  For each representative $U$ of a class of conjugated subgroups 
$\widetilde{U}$ left coset representatives have to be chosen and fixed.
The index $|\mathrm{D}_N|/|\mathrm{D}_l|=\lambda$ and the rotations
$C_N^j$, $j=0,1,\dots,\lambda-1$ will play this role. In the case of 
cyclic subgroups $\mathrm{C}_k$ one needs additional $\lambda$ 
representatives. They can be chosen as $U_j$, $j=0,1,\dots,\lambda-1$. 
The choice of coset representatives is fixed in the further considerations.

Considering double cosets $V\backslash\mathrm{D}_N/U$ we determine at first 
their number in different cases using results presented in Sec.~3 of the 
previous work \cite{prev}.  If both $U$ and $V$ are cyclic then only 
classes containing rotations $C_N^j$ have to be taken into account and one 
obtains 
 $$
    |\mathrm{C}_k\backslash \mathrm{D}_N/\mathrm{C}_l|= 
       2\gcd\,(\kappa,\lambda)=2\chi\,,
 $$
 where `gcd' denotes the greatest common divisor and $\kappa=N/k, 
\lambda=N/l$ are co-divisors of $k$ and $l$, respectively. Note that for 
mutually prime co-divisors $\kappa,\lambda$ there are two double cosets 
only: one contains rotations $C_N^j$, whereas the second --- elements 
$U_j$. If $V=\mathrm{D}_k$ (or $U=\mathrm{D}_l$ and $V=\mathrm{C}_k$), then 
the only change is the order of one subgroup, so the result is
 $$
    |\mathrm{C}_k\backslash \mathrm{D}_N/\mathrm{D}_l|= 
      |\mathrm{D}_k\backslash \mathrm{D}_N/\mathrm{C}_l|= \chi\,.
 $$

 If both subgroups are dihedral then the cases of odd and even $N$ have to 
be considered separately. Moreover, in the case of even $N$ a dihedral 
subgroup $\mathrm{D}_k$ can be in one of three types: $\mathrm{D}_k^0$, 
$\mathrm{D}_k^1$ or $\mathrm{D}_k$. In the first case it contains elements
$U_j$ with even indices $j$, in the second case all $j$'s are odd and
in the last case odd and even indices $j$ are present. The final results
are as follows:
 \begin{itemize}
  \item $N$ odd
    $$
      |\mathrm{D}_k\backslash \mathrm{D}_N/\mathrm{D}_l| = (\chi+1)/2\,;
    $$
  \item $N$ even
     \begin{eqnarray*}
       |\mathrm{D}_k^0\backslash \mathrm{D}_N/\mathrm{D}_l^0| = 
          |\mathrm{D}_k^1\backslash \mathrm{D}_N/\mathrm{D}_l^1| &= &
          (\chi+2)/2\,;\\
       |\mathrm{D}_k^0\backslash \mathrm{D}_N/\mathrm{D}_l^1| &= &
           \chi/2\,;\\
       |\mathrm{D}_k^0\backslash \mathrm{D}_N/\mathrm{D}_l| = 
          |\mathrm{D}_k^1\backslash \mathrm{D}_N/\mathrm{D}_l| = 
          |\mathrm{D}_k\backslash \mathrm{D}_N/\mathrm{D}_l| &= &
          (\chi+1)/2\,.
      \end{eqnarray*}
 \end{itemize} 

The above presented formulas are symmetric, {\it i.e.}\ the subgroups
$U$ and $V$ can be transposed. However, in a general case a set of double
cosets $V\backslash G/U$ is not symmetric, {\it i.e.}\ it differs from
the set $U\backslash G/V$ (they have the same orders only). The same can
be said about representatives $g_d$ of double cosets. Therefore, both 
cases should be considered separately and for each pair $V,U$ we obtain
two sets of double coset representatives. On the other hand, the operators
considered are Hermitian, so $h_{U\mu r,V\nu s}=h^*_{V\nu s,U\mu r}$ and, 
therefore, it is enough to determine one element of each pair. 
Nevertheless, for the sake of completeness all cases are considered below. 

 Since the cyclic subgroups are self-conjugated, then for a given $V$ 
all double cosets $Vg\mathrm{C}_l$ have the same order (see Eq.~(13) in 
\cite{prev}):
 $$
  |Vg\mathrm{C}_l|=|\mathrm{C}_lgV|= \frac{l|V|}{|V\cap\mathrm{C}_l|}\,. 
 $$
 The stabilizer in the denominator for $V=\mathrm{D}_k$ or $\mathrm{C}_k$
is equal to $\mathrm{C}_{\gcd(k,l)}$. Therefore,
 $$
  |\mathrm{C}_kg\mathrm{D}_l| = |\mathrm{D}_kg\mathrm{C}_l| =
    2|\mathrm{C}_kg\mathrm{C}_l| = 2\frac{kl}{\gcd(k,l)}=
      2\,\mathrm{lcm}(k,l)\,, 
 $$
 where `lcm' denotes the least common multiplicity. This result agrees with
 the above presented formulas, since $\chi=\gcd(\kappa,\lambda)=N/h$ with 
$h=\mathrm{lcm}(k,l)$. These double cosets are represented by $g_d=C_N^j$
with $j=0,1,\dots,\chi-1$ if one subgroup is cyclic. If both subgroups are 
cyclic, then the next $\chi$ representatives are $g_d=U_j$, $0\le j<\chi$. 
The cardinality of double cosets just obtained does not depend on the 
representative $g$, so Eq.~(\ref{Hgd}) can be simplified if one of 
subgroups is cyclic. To shorten the notation let us introduce two symbols: 
(i) the product $B^{UV\Gamma}_{rs}(g_d)\langle U\mu g_d|\mathcal{H}| 
 V\nu E\rangle$ is denoted as $f(g_d)$ ($U$, $V$, $\Gamma$ {\it etc.}\ 
are omitted, but they can be easily determined from the context) and (ii)
$\mathrm{sd}(k,l)=\mathrm{lcm}(k,l)/\gcd(k,l)$.\footnote{The symbol `sd'
means the {\em symmetric division} and has been introduced by analogy to 
the symmetric difference of sets: $\mathrm{sd}(k,l)$ is a product of primes, 
which appear in decomposition of either $k$ or $l$; the authors have not 
found any special name for this symbol in available books on the number
theory.} Using these symbols one can write
 \begin{eqnarray}
   h_{\mathrm{C}_l \mu r,\mathrm{C}_k \nu s}(\Gamma)&=&
    \mathrm{sd}(k,l)^{1/2} \sum_{d=0}^{\chi-1} f(C_N^d) + f(U^d)\,,
     \label{Hgdcc} \\
   h_{\mathrm{D}_l \mu r,\mathrm{C}_k \nu s}(\Gamma) =
     h_{\mathrm{C}_l \mu r,\mathrm{D}_k \nu s}(\Gamma)&=&
       (2\mathrm{sd}(k,l))^{1/2}\: \sum_{d=0}^{\chi-1} f(C_N^d) \,.
     \label{Hgdcd} 
 \end{eqnarray}
 
The most complicated is the case when both subgroups $U$ and $V$ are
dihedral, because double cosets obtained may have different lengths and
there are three different values of the number of double cosets. In fact, 
we are not interested in their detailed structure, but we have to know 
the number of representatives $g$ of left cosets $gU$ contained in a 
double coset $Vg_dU$. The appropriate rules can be derived if one notices 
that the action of $V=\mathrm{D}_k$ on the set of left cosets 
$\mathrm{D}_N/\mathrm{D}_l$ can be performed in two steps: (i) the action
of the cyclic subgroup $\mathrm{C}_k\subset\mathrm{D}_k$ considered above
and (ii) the action of the generator $U_0$ or $U_1$, depending on the 
type of $V$. The other elements $U_{q+p\kappa}$, $q=0,1$, $0<p<k$, do not
yield new results since $U_{q+p\kappa} =U_qC_N^{(k-p)\kappa}$ and
the action of $C_N^{(k-p)\kappa}$ has been already taken into account.
After the first step we have at hand $\chi$ orbits represented by $C_N^j$, 
$0\le j<\chi$. Each coset $\mathrm{C}_kC_N^j\mathrm{D}_l$ contains also
$\lambda'=\lambda/\chi$ representatives of left cosets $g\mathrm{D}_l$ 
in the form $C_N^{r\chi+j}$, $0\le r<\lambda'$. From the multiplication 
rule (\ref{mulrul}) we obtain ($h=N/\chi$): 
 \begin{eqnarray*}
   U_0C_N^j=C_N^{\chi-j}U_{\chi(h-1)}&\quad\mbox{or}\quad&
     U_0C_N^j =C_N^{\chi-j-1}U_{\chi(h-1)+1}\,;\\
   U_1C_N^j=C_N^{\chi-j+1}U_{\chi(h-1)}&\mbox{or}&
    U_1C_N^j =C_N^{\chi-j}U_{\chi(h-1)+1}\,.
 \end{eqnarray*}
 The following four cases have to be discussed:
\begin{itemize}
  \item  $U=\mathrm{D}_l^0$, $V=\mathrm{D}_k^0$: It follows from the first
    formula that cosets $\mathrm{C}_k C_N^j\mathrm{D}_l^0$ and
    $\mathrm{C}_k C_N^{\chi-j}\mathrm{D}_l^0$ are merged. There is one
    fixed point, namely $j=0$.  If $\chi$ is even then there is another 
    fixed point $j=\chi/2$.
  \item  $U=\mathrm{D}_l^1$, $V=\mathrm{D}_k^1$: In this case the last 
     formula should be used so the results are the same as above.
  \item  $U=\mathrm{D}_l^1$, $V=\mathrm{D}_k^0$: Orbits 
     $\mathrm{C}_kC_N^j\mathrm{D}_l^1$ and 
     $\mathrm{C}_kC_N^{j'}\mathrm{D}_l^1$ are merged if 
     $j'=-(1+j)\pmod{\chi}$. The fixed point is obtained for 
     $2j=-1\pmod{\chi}$, so it is possible for odd $\chi$ only and in this 
     case $j=(\chi-1)/2$. 
  \item  $U=\mathrm{D}_l^0$, $V=\mathrm{D}_k^1$: In this case the 
  congruence $j'=1-j\pmod{\chi}$ has to be satisfied to merge orbits and 
  the fixed point is obtained for $j=(\chi+1)/2$, if $\chi$ is odd. In 
  this case it may be more convenient to use $j=1,2,\dots,\chi/2$ for 
  $\chi$ even and $j=1,2,\dots,(\chi-1)/2,(\chi+1)/2$ for $\chi$ odd. 
\end{itemize}
  In Table~\ref{dcd6} all possible cases are presented for $N=6$ and 
$U,V=\mathrm{D}_1^0,\mathrm{D}_1^1$, and $\mathrm{D}_2$. Each entry is a 
set of representatives $g$ of left cosets $gU$, which are contained
in the same double coset $Vg_dU$. 

\begin{table}
  \caption{Representatives of cosets $gU$ contained in the same double
     coset $VgU$ for the dihedral group $\mathrm{D}_6$ and $U,V=
     \mathrm{D}_1^0,\mathrm{D}_1^1,\mathrm{D}_2$; underbraced elements 
     belong to the same orbit under the action of the cyclic subgroup 
     $\mathrm{C}_2$ of $V=\mathrm{D_2}$}\label{dcd6} 
 \begin{center}
  \begin{tabular}{c|ccc}
      ~~~$U$ & $\mathrm{D}_1^0$ & $\mathrm{D}_1^1$ & $\mathrm{D}_2$ \\ 
   $V$~~~ & && \\\hline\hline
   $\mathrm{D}_1^0$ & $\{E\}$ & $\{E,C_6^5\}$ & $\{E\}$ \\
    & $\{C_6,C_6^5\}$ & $\{C_6,C_6^4\}$ & $\{C_6,C_6^2\}$ \\
    & $\{C_6^2,C_6^4\}$ & $\{C_6^2,C_6^3\}$ & \\
    & $\{C_6^3\}$ & & \\ \hline
   $\mathrm{D}_1^1$ 
    & $\{E,C_6\}$ & $\{E\}$ & $\{E,C_6\}$ \\
    & $\{C_6^2,C_6^5\}$ & $\{C_6,C_6^5\}$ & $\{C_6^2\}$ \\
    & $\{C_6^3,C_6^4\}$ & $\{C_6^2,C_6^4\}$ & \\
    & & $\{C_6^3\}$ & \\ \hline
   $\mathrm{D}_2$ 
    & $\{\underbrace{E,C_6^3\}}{}$ 
    & $\{\underbrace{E,C_6^3,}{}\underbrace{C_6^2,C_6^5\}{}}$ & $\{E\}$ \\
    & $\{\underbrace{C_6,C_6^4}{},\underbrace{C_6^2,C_6^5}{}\}$ 
    & $\{\underbrace{C_6,C_6^4}{}\}$ & $\{C_6,C_6^2\}$ 
  \end{tabular}
 \end{center}
\end{table}

Taking into account the results presented above one obtains the following
formulas for matrix elements:
 \begin{eqnarray}
  h_{\mathrm{D}_l^0\mu r,\mathrm{D}_k^0\nu s}(\Gamma) =
    h_{\mathrm{D}_l^1\mu r,\mathrm{D}_k^1\nu s}(\Gamma) &=&
      \mathrm{sd}(k,l)^{1/2}\left(f(E) + 2\sum_{d=1}^{\chi/2-1} f(C_N^d)
       + f(C_N^{\chi/2})\right); \label{h1} \\
  h_{\mathrm{D}_l^0\mu r,\mathrm{D}_k^1\nu s}(\Gamma) &=&
    2\,\mathrm{sd}(k,l)^{1/2}\sum_{d=1}^{\chi/2} f(C_N^d)\,; \label{h2} \\
  h_{\mathrm{D}_l^0\mu r,\mathrm{D}_k\nu s}(\Gamma) =
   h_{\mathrm{D}_l\mu r,\mathrm{D}_k^0\nu s}(\Gamma) &=& 
   h_{\mathrm{D}_l\mu r,\mathrm{D}_k\nu s}(\Gamma)  \nonumber  \\
   &=&     \mathrm{sd}(k,l)^{1/2}\left( f(E)
    +2\sum_{d=1}^{(\chi-1)/2} f(C_N^d)\right)\,; \label{h3} \\
 h_{\mathrm{D}_l^1\mu r,\mathrm{D}_k^0\nu s}(\Gamma) &=&
   2\,\mathrm{sd}(k,l)^{1/2}\sum_{d=0}^{\chi/2-1} f(C_N^d)\,; \label{h4} \\
 h_{\mathrm{D}_l^1\mu r,\mathrm{D}_k\nu s}(\Gamma) &=&
   \mathrm{sd}(k,l)^{1/2}\left(2\sum_{d=0}^{(\chi-3)/2} f(C_N^d)
   +f(C_N^{(\chi-1)/2})\right)\,; \label{h5} \\
 h_{\mathrm{D}_l\mu r,\mathrm{D}_k^1\nu s}(\Gamma) &=&
    \mathrm{sd}(k,l)^{1/2}\left(2\sum_{d=1}^{(\chi-1)/2} f(C_N^d)
     +f(C_N^{(\chi+1)/2})\right)\,. \label{h6}
 \end{eqnarray}

\subsection{Transitive irreps}
  Let us start with the cyclic subgroups $\mathrm{C}_l= 
\langle C_N^\lambda\rangle$ with $\lambda=N/l$. Since for all $j=0,1,2,N-1$ 
we have $\chi^{A_{1,2}}(C_N^j)=1$ (see Eqs.~(\ref{a1rep},\ref{a2rep})), then
 \begin{equation}\label{tra1a2c}
   n(A_1,R^{\mathrm{D}_N:\mathrm{C}_l}) 
    =n(A_2,R^{\mathrm{D}_N:\mathrm{C}_l})= 1\,.
 \end{equation}
 In the case of $N$ even there is also another pair of one-dimensional
irreps, namely $B_1$ and $B_2$. It can be shown that both of them appears
in the decomposition of $R^{\mathrm{D}_N:\mathrm{C}_l}$ if and only if 
$\lambda$ is even. The two-dimensional irreps $E_p$ always appear twice,
{\it i.e.}\ $n(E_p,R^{\mathrm{D}_N:\mathrm{C}_l})=2$ if $\gcd(p,l)=l$. Since
$p<N/2$ then only subgroups $\mathrm{C}_l$ with $l<N/2$ lead to decompositions
containing $E_p$. 

 The dihedral groups $\mathrm{D}_l$ have twice more elements then
$\mathrm{C}_l$ then the dimensions of $R^{\mathrm{D}_N:\mathrm{D}_l}$ are
two times smaller. At first, one can notice that the irrep $A_2$ does not
appear in the appropriate decompositions. The case $\Gamma=E_p$ is simply,
since one obtains
 $$
  n(E_p,R^{\mathrm{D}_N:\mathrm{D}_l})
    =n(E_p,R^{\mathrm{D}_N:\mathrm{C}_l})/2 =1 \quad 
	  \mathrm{if}\quad  \gcd(p,l)=l\,. 
 $$
 The irreps $B_1$ and $B_2$ (admissible for $N$ even only) appears depending
on the type of the group $\mathrm{D}_l$: 
 $$
  n(B_1,R^{\mathrm{D}_N:\mathrm{D}_l^0})
    = n(B_2,R^{\mathrm{D}_N:\mathrm{D}_l^1}) =1\,. 
 $$
 The results obtained are illustrated in Table~\ref{class} for $N=6$. 

\begin{table}
  \caption{Decompositions of transitive representations into irreps for 
    $\mathrm{D}_6$ (subgroups are determined by their generators)}
    \label{class}
  \begin{center}
    \begin{tabular}{ll}
    $U$  & $R^{\mathrm{D}_6:U}$\\ \hline
     $\mathrm{C}_1=\{E\}$ & 
          $A_1\oplus A_2\oplus B_1\oplus B_2\oplus 2E_1\oplus 2E_2$\\
     $\mathrm{D}_1^0=\langle U_0\rangle$ 
         & $A_1\oplus B_1\oplus E_1\oplus E_2$ \\
    $\mathrm{D}_1^1=\langle U_1\rangle$ 
       & $A_1\oplus B_2\oplus E_1\oplus E_2$ \\
    $\mathrm{C}_2=\langle C_6^3\rangle$ & $A_1\oplus A_2\oplus 2E_2$ \\
    $\mathrm{D}_2=\langle C_6^3,U_0\rangle$ & $A_1\oplus E_2$ \\
    $\mathrm{C}_3=\langle C_6^2\rangle$ 
	   & $A_1\oplus A_2\oplus B_1\oplus B_2$ \\
    $\mathrm{D}_3^0=\langle C_6^2,U_0\rangle$ & $A_1\oplus B_1$ \\
    $\mathrm{D}_3^1=\langle C_6^2,U_1\rangle$ & $A_1\oplus B_2$ \\
    $\mathrm{C}_6=\langle C_6\rangle$ & $A_1\oplus A_2$ \\
    $\mathrm{D}_6=\langle C_6,U_0\rangle$ & $A_1$
  \end{tabular}
 \end{center}
\end{table}
 
\subsection{Reduction coefficients}

 A general formula for the reduction coefficients, see Sec.~2.4 in 
\cite{prev}, yields that one-dimensional irreps $\Gamma$ always lead
to trivial reduction coefficients $A^{U\Gamma}=1$. The indices $\gamma=1$
and $r=1$ can be omitted. So only the case $\Gamma=E_p$ has to be considered
in more details. The vector of irreducible basis related with the reduction 
$E_p\downarrow U$ in this case can be written as
 $$
  |UE_p r\rangle=A^{UE_p}_{x r} |E_px\rangle+ A^{UE_p}_{y r} |E_py\rangle\,, 
 $$
 where $r=1$ for the dihedral ($\mathrm{D}_l$) and $r=1,2$ for the cyclic 
($\mathrm{C}_l$) subgroups $U$, respectively. In the first case we have to 
distinguish  subgroups $\mathrm{D}_l^0$ and $\mathrm{D}_l^1$. The standard
choice of the basis $\{|E_px\rangle,|E_py\rangle\}$ and Eq.~(\ref{repemat})
yield the matrices:
 $$
   E_p(U_0)=\left(\begin{array}{rr} 1 & 0 \\ 0 & -1\end{array}\right)\,,
     \qquad E_p(U_1)=\left(\begin{array}{rr} 
        \cos(2\pi p/N) &  \sin(2\pi p/N) \\  
        \sin(2\pi p/N) &  -\cos(2\pi p/N)  \end{array}\right)\,.
 $$
 Therefore, one obtains (for representations $E_p$ appearing in the 
decompositions of $R^{\mathrm{D}:\mathrm{D}_l}$, {\it i.e.}\ for 
$\gcd(p,l)=l$):
   \begin{eqnarray}
      A^{\mathrm{D}_l^0\,E_p}_x=1\,,&\qquad&
      A^{\mathrm{D}_l^0\,E_p}_y=0\,;\\
      A^{\mathrm{D}_l^1\,E_p}_x=\cos(\varphi/2)\,,&\qquad&
      A^{\mathrm{D}_l^1\,E_p}_y=\sin(\varphi/2)\,,
   \end{eqnarray}
 where $\varphi=2\pi p/N=2\pi p_l/\lambda$ with $p_l=p/l$. The cyclic 
subgroups yield $r=1,2$ and it can be shown that there are no
restrictions imposed on the coefficients $A^{\mathrm{C}_l\,E_p}_{\gamma r}$,
as long as they determine orthonormal vectors $|\mathrm{C}_lE_p1\rangle$, 
$|\mathrm{C}_lE_p2\rangle$. The simplest choice is to put
   \begin{eqnarray}
      A^{\mathrm{C}_l\,E_p}_{x1}=1\,,&\qquad&
      A^{\mathrm{C}_l\,E_p}_{y1}=0\,;\\
      A^{\mathrm{C}_l\,E_p}_{x2}=0\,,&\qquad&
      A^{\mathrm{C}_l\,E_p}_{y2}=1\,.
   \end{eqnarray}

Now we have to calculate the coefficients $B^{UV\Gamma}_{rs}(g_j)$ for the
representatives of left cosets $g_jU$. Let us start from one-dimensional
irreps $\Gamma=A_1,A_2,B_1,B_2$. Since the coefficients $A^{U\Gamma}=1$
and $r,s=1$, then 
 \begin{equation} 
  B^{UV\Gamma}(g_j)= \Gamma(g_j) =\chi^{\Gamma}(g_j) \,.
 \end{equation}
 Characters of these representations are given by 
Eqs.~(\ref{a1rep})--(\ref{b2rep}). In particular, the unit irrep $A_1$ 
always yields $B^{UVA_1}(g_j)=1$. 

The two-dimensional irreps $E_p$ appear in decompositions of 
$R^{\mathrm{D}_N:\mathrm{C}_l}$ and $R^{\mathrm{D}_N:\mathrm{D}_l}$ for
$l<N/2$ and $p=lp_l$; if $U$ or $V$ is the cyclic subgroup then $E_p$ 
appears twice. Moreover, there are two types of the dihedral
subgroups for $N$ even. It leads to 16 different cases of coefficients
$B^{UV\,E_p}_{rs}$, which will be presented below. We are interested
in the coefficients $B^{UV\,E_p}_{rs}$ when the irrep $E_p$ appears in the 
decompositions of $R^{\mathrm{D}_N:U}$ and $R^{\mathrm{D}_N:V}$. Let,
as in Sec.~\ref{cosets}, $V=\mathrm{C}_k$ or $\mathrm{D}_k$ and
$U=\mathrm{C}_l$ or $\mathrm{D}_l$. Then the index $p$ of the irrep $E_p$
has to satisfy both conditions: $p=p_kk$ and $p=p_ll$. Therefore,
$p=\mathrm{lcm}(k,l)p'=hp'=Np'/\chi$ and $\varphi=2\pi p/N=2\pi p'/\chi$.

Let subgroups $U,V$ be cyclic. Then both repetition indices $r,s=1,2$, so 
there are four different coefficients $B^{UV\,E_p}_{rs}(g_j)$. We have
chosen elements $C_N^j$ and $U_j$, $j=0,1,\dots,\lambda-1$ as representatives 
of the left cosets $g_j\mathrm{C}_l$ with representation matrices  
  $$
    E_p(C_N^j)=\left(\begin{array}{rr} 
           \cos j\varphi  &  -\sin j\varphi  \\  
         \sin j\varphi  &   \cos j\varphi   \end{array}\right)\,,\qquad
    E_p(U_j)=\left(\begin{array}{rr} 
      \cos j\varphi  &   \sin j\varphi  \\  
       \sin j\varphi  &  -\cos j\varphi   \end{array}\right)\,.
  $$
Taking all these relations into account one obtains
\begin{eqnarray}
 B^{\mathrm{C}_l\mathrm{C}_k\,E_p}_{11}(C_N^j)=
  B^{\mathrm{C}_l\mathrm{C}_k\,E_p}_{22}(C_N^j)=
  B^{\mathrm{C}_l\mathrm{C}_k\,E_p}_{11}(U_j)=
  -B^{\mathrm{C}_l\mathrm{C}_k\,E_p}_{22}(U_j)&=& \cos j\varphi\,,
       \label{CCcos}\\
 B^{\mathrm{C}_l\mathrm{C}_k\,E_p}_{12}(C_N^j)=  
  -B^{\mathrm{C}_l\mathrm{C}_k\,E_p}_{21}(C_N^j)= 
   B^{\mathrm{C}_l\mathrm{C}_k\,E_p}_{12}(U_j)=   
   B^{\mathrm{C}_l\mathrm{C}_k\,E_p}_{21}(U_j)&=& \sin j\varphi\,.
     \label{CCsin}
\end{eqnarray}
When the subgroup $V$ is dihedral than the repetition index can be omitted,
but two types of dihedral subgroups have to be considered. It leads to
the following formulas:
 \begin{eqnarray} 
  B^{\mathrm{C}_l\mathrm{D}_k^0E_p}_{r=1}(C_N^j)= 
  B^{\mathrm{C}_l\mathrm{D}_k^0E_p}_{r=1}(U_j)&=&
\cos j\varphi \,, \label{CD0cos} \\
  B^{\mathrm{C}_l\mathrm{D}_k^0E_p}_{r=2}(U_j)=
  -B^{\mathrm{C}_l\mathrm{D}_k^0E_p}_{r=2}(C_N^j)&=&
  \sin j\varphi  \,, \label{CD0sin} \\
  B^{\mathrm{C}_l\mathrm{D}_k^1E_p}_{r=1}(C_N^j)=
  B^{\mathrm{C}_l\mathrm{D}_k^1E_p}_{r=1}(U_j)&=&
  \cos (j-1/2)\varphi   \,, \label{CD1cos} \\
  B^{\mathrm{C}_l\mathrm{D}_k^1E_p}_{r=2}(U_j)=
  -B^{\mathrm{C}_l\mathrm{D}_k^1E_p}_{r=2}(C_N^j)&=&
  \sin (j-1/2)\varphi  \,.  \label{CD1sin}
 \end{eqnarray} 
The case $U=\mathrm{D}_l$, $V=\mathrm{C}_k$ is a bit easier to discuss,
since the left cosets $g_j\mathrm{D}_l$ are represented by the  
elements $C_N^j$ only,
 \begin{eqnarray} 
  B^{\mathrm{D}_l^0\mathrm{C}_kE_p}_{s=1}(C_N^j)&=&
  \cos j\varphi  \,,\label{D0Cc}\\
  B^{\mathrm{D}_l^0\mathrm{C}_kE_p}_{s=2}(g_j)&=&
 \sin j\varphi  \,,\label{D0Cs}\\
  B^{\mathrm{D}_l^1\mathrm{C}_kE_p}_{s=1}(C_N^j)&=&
  \cos (j+1/2)\varphi  \,,\label{D1Cc}\\
  B^{\mathrm{D}_l^1\mathrm{C}_kE_p}_{s=2}(g_j)&=&
 \sin (j+1/2) \varphi \,.\label{D1Cs}  
 \end{eqnarray} 
In the last four cases both subgroups are dihedral, so $g_j=C_N^j$ and 
different types of subgroups are considered. The final results are as
follows:
 \begin{eqnarray} 
  B^{\mathrm{D}_l^0\mathrm{D}_k^0E_p}(C_N^j)&=&
  \cos j\varphi\,,\label{D00c}  \\
  B^{\mathrm{D}_l^1\mathrm{D}_k^0E_p}(C_N^j)&=&
  \cos (j+1/2)\varphi  \,,\label{D10c}\\
  B^{\mathrm{D}_l^0\mathrm{D}_k^1E_p}(C_N^j)&=&
  \cos (j-1/2)\varphi  \,,\label{D01c}\\
  B^{\mathrm{D}_l^1\mathrm{D}_k^1E_p}(C_N^j)&=&
  \cos j\varphi  \,.\label{D11c}
 \end{eqnarray} 

The equation presented together with Eqs.~(\ref{Hgdcc}--\ref{h6})
enable us to determine matrix elements for any operator commuting 
with $P(g)$.

\section{Example}\label{example}
Let us consider a ring of six spins $s=1$ with the symmetry group
$\mathrm{D}_6$ and the antiferromagnetic interactions of the 
nearest-neighbors ($J=-1$)
 $$
   \mathcal{H}=\sum_{j=1}^6 \mathbf{s}_j\cdot\mathbf{s}_{j+1}\,,\qquad 
  6+j\equiv j\,;
 $$
 The dimension of the space of states is $(2s+1)^N=729$, so it is not so 
large, but it is sufficient to illustrate application of the method 
presented in this paper. 

 Since in this case the magnetization is a good quantum number, then 
at first we consider ordered and non-ordered partitions of $N=6$
into no more than $2s+1=3$ parts. There are seven ordered partitions:
$[600]$, $[510]$, $[420]$, $[411]$, $[330]$, $[321]$, and $[222]$.  
They lead to 28 non-ordered partitions, which were presented in the 
previous paper \cite{prev} in Table~1. On the other hand, some simple
combinatorial considerations yield the dimensions of subspaces with
given magnetization $M$ \cite{prev,oldwsf}. The results obtained are
collected in Table~\ref{Mandp}. Due to the time-reversal symmetry it is
enough to consider the states with $M\ge0$. 

\begin{table}\label{Mandp}
\caption{Dimensions of subspaces of given magnetization $M\ge0$ and
non-ordered partitions related with them}
\begin{center}
\begin{tabular}{ccrr}
$M$ & non-ordered partitions $[k]$ & 
  \multicolumn{1}{c}{$|\mathcal{B}_{[k]}|$} & 
  \multicolumn{1}{c}{$\dim L_M$} \\ \hline
0 & [303], [222], [141], [060] & 20, 90, 30, 1 & 141 \\
1 & [213], [132], [051] & 60, 60, 6 & 126 \\
2 & [204], [123], [042] & 15, 60, 15 &  90 \\ 
3 & [114], [033] & 30, 20 &  50 \\
4 & [105], [024] &  6, 15  &  21 \\ 
5 & [015] &  6 &   6 \\
6 & [006] &  1 &   1 
\end{tabular} 
 \end{center}
\end{table}

The non-ordered partitions determine orbits of the symmetric group
$\Sigma_6$. The action of $\mathrm{D}_6\subset\Sigma_6$ decomposes them
into orbits labeled by subgroups $U\subset\mathrm{D}_6$ (see Sec.~\ref{subgr}
and Fig.~\ref{ld6}). These decompositions depend only on the type of 
$[k]$, {\it i.e.} on the ordered partition $[\kappa]$, and are 
presented below (the number before a subgroup symbol $U$ denotes the
multiplicity of orbits of the same type $\widetilde{U}$):
 \begin{eqnarray*}
&&[6]\,\colon {\mathrm D}_6;\quad [51]\colon {\mathrm D}_1^0;\quad
[42]\,\colon {\mathrm D}_2, {\mathrm D}_1^0, {\mathrm D}_1^1;\quad
[33]\,\colon {\mathrm D}_3^0, {\mathrm D}_1^0, {\mathrm C}_1;\\
&& [411]\,\colon {\mathrm D}_1^0, 2{\mathrm C}_1;\quad  
  [321]\,\colon  2{\mathrm D}_1^0, 4{\mathrm C}_1;\quad 
 [222]\,\colon {\mathrm C}_2, 3{\mathrm D}_1^0, 3{\mathrm D}_1^1, 
 4{\mathrm C}_1 \,.
 \end{eqnarray*}
These results yield classification of orbits by the magnetization $M$
and the type $\widetilde{U}$:
 $$
  M=6\; \colon\; \mathrm{D}_6\,; \quad 
     M=5\; \colon\; \mathrm{D}_1^0\,; \quad 
     M=4\; \colon\; \mathrm{D}_2, 2\mathrm{D}_1^0, \mathrm{D}_1^1\,; \quad
   M=3\;\colon\; \mathrm{D}_3^0, 2\mathrm{D}_1^0, 3\mathrm{C}_1\,;
$$$$ 
 M=2\; \colon\; 2\mathrm{D}_2, 4\mathrm{D}_1^0, 2\mathrm{D}_1^1, 
        4\mathrm{C}_1\,; \quad
   M=1\; \colon\; 5\mathrm{D}_1^0, 8\mathrm{C}_1\,; \quad
   M=0\;\colon\; \mathrm{D}_6, \mathrm{D}_3^0, \mathrm{C}_2, 
     5\mathrm{D}_1^0, 3\mathrm{D}_1^1, 7\mathrm{C}_1\,. 
 $$
  Now, 
taking into account the decompositions presented in Table~\ref{class},
we can write down the multiplicities $n(\Gamma,P\downarrow L_M)$ for 
each $M\ge0$. The numbers presented in Table~\ref{multi} give us
the dimension of the eigenproblem $\mathcal{H}|\psi\rangle=E|\psi\rangle$
restricted to the states with a given magnetization $M$ and the symmetry 
described by $\Gamma$. The largest dimension is reached for $M=0$ and 
$\Gamma=E_2$. However, when we are interested in the ground state only,
then $\Gamma=A_1$ and the largest eigenproblem has dimension $18\approx
729/40$. We can decrease these numbers even more taking into the square
of the total spin $\mathbf{S}^2$. Since eigenvalues of this operator
are $S(S+1)$ for $0\le S\le 6$, then it is enough to find the
corresponding eigenvectors and express the Hamiltonian matrix in this
new basis. Table \ref{multi} provides us also with dimensions 
$\dim L_{M=S,\Gamma}$ of subspaces with a given total spin $S$ and the 
maximum magnetization $M=S$ corresponding to an irrep $\Gamma$. It is enough 
to subtract adjoining rows of this table (except for the first one since 
$\dim L_{M=S=Ns}=\dim L_{M=Ns}$), because sates with $M=S$ are these
with a given $M$ except for those with $S=M+1$, which are listed
in the previous row. For example, the dimensions $\dim L_{M=S,A_1}$ are 
5, 1, 6, 2, 3, 0, and 1 for $S=0,1,\dots,6$, respectively. In the same
way one obtains that the subspace of states with $M=S=0$ has the dimension 
$\dim L_{M=S=0}=15$ and 
 $$
   P\downarrow L_{M=S=0} = 5 A_1\oplus 2B_2 \oplus E_1\oplus 3E_2\,,
 $$
 so the dimension of the Hamiltonian eigenproblem in this case is
at most five. The overall maximum is reached for 
$M=S=2$ and $\Gamma=E_1$ and equals eight, {\it i.e.}\ about one
ninetieth of the original dimension $\dim L=729$.

\begin{table}
\caption{Classification of states by the magnetization $M\ge0$ and the 
irrep $\Gamma$} \label{multi}
\begin{center}
\begin{tabular}{c*{7}{r}}
$M$ & \multicolumn{1}{c}{$A_1$} & \multicolumn{1}{c}{$A_2$} 
    & \multicolumn{1}{c}{$B_1$} & \multicolumn{1}{c}{$B_2$} 
    & \multicolumn{1}{c}{$E_1$} & \multicolumn{1}{c}{$E_2$} 
    & $\dim L_M$ \\ \hline
 6  &  1 & 0 &  0 &  0 &  0 &  0 &   1 \\ 
 5  &  1 & 0 &  1 &  0 &  1 &  1 &   6 \\ 
 4  &  4 & 0 &  2 &  1 &  4 &  3 &  21 \\ 
 3  &  6 & 3 &  6 &  3 &  8 &  8 &  50 \\ 
 2  & 12 & 4 &  8 &  6 & 16 & 14 &  90 \\ 
 1  & 13 & 8 & 13 &  8 & 21 & 21 & 126 \\ 
 0  & 18 & 8 & 13 & 10 & 22 & 24 & 141 \\ \hline
Total & 92 & 38 & 73 & 46 & 122 & 118 & 
\end{tabular}
 \end{center}
\end{table}

\begin{figure}
  \begin{center}\epsfxsize 11cm \epsfbox{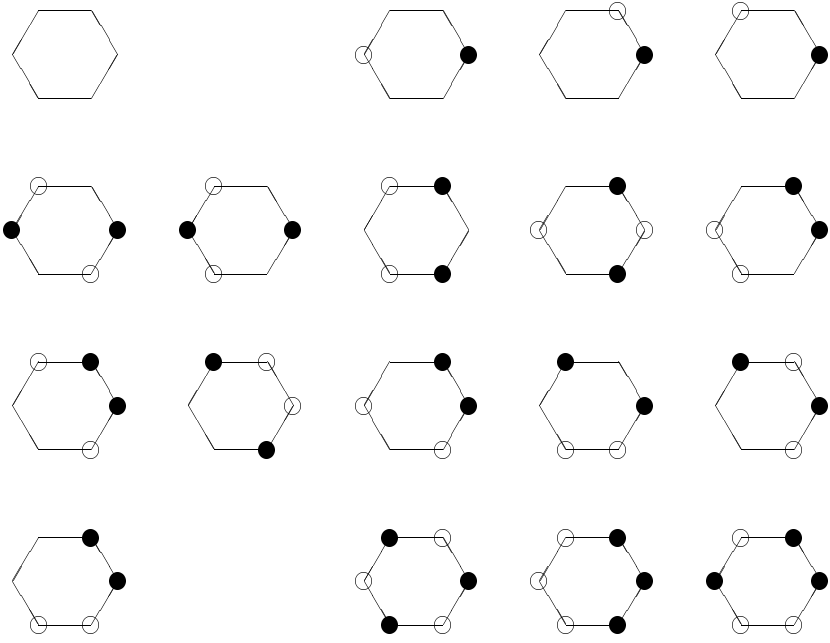} \end{center}
\caption{Orbit representatives for $N=6$, $s=1$ and $M=0$; full and empty
circles denote $m_j=+1,-1$, respectively}
\label{orep}
\end{figure}

Let us restrict the further considerations to the case $M=S=0$ and
$\Gamma=A_1$. There are 18 orbits grouped into four classes labeled by the 
non-ordered partitions [060], [141], [222], and [303]; their representatives 
are presented in Fig.~\ref{orep}.  Using the symbols $|U [k] \mu\rangle$, 
where $U$ is a stabilizer, $[k]$ --- a non-ordered partition, and $\mu$ --- a 
configuration expressed as a series $(m_1,\dots,m_6)$ with $\pm$ standing for 
$\pm1$, respectively, these representatives can be written as:
 $$
 \begin{array}{r@{\;=\;}l@{\qquad}r@{\;=\;}l}
  | 1\rangle & |\mathrm{D}_6   \,[060]\,(0,0,0,0,0,0)\rangle\,, & 
    | 2\rangle & |\mathrm{D}_1^0 \,[141]\,(+,0,0,-,0,0)\rangle\,, \\
  | 3\rangle & |\mathrm{C}_1   \,[141]\,(+,-,0,0,0,0)\rangle\,, & 
    | 4\rangle & |\mathrm{C}_1   \,[141]\,(+,0,-,0,0,0)\rangle\,, \\
  | 5\rangle & |\mathrm{C}_2   \,[222]\,(+,0,-,+,0,-)\rangle\,, & 
    | 6\rangle & |\mathrm{D}_1^0 \,[222]\,(+,0,-,+,-,0)\rangle\,, \\
  | 7\rangle & |\mathrm{D}_1^0 \,[222]\,(0,+,-,0,-,+)\rangle\,, & 
    | 8\rangle & |\mathrm{D}_1^0 \,[222]\,(-,+,0,-,0,+)\rangle\,, \\
  | 9\rangle & |\mathrm{D}_1^1 \,[222]\,(+,+,0,-,-,0)\rangle\,, & 
    |10\rangle & |\mathrm{D}_1^1 \,[222]\,(+,+,-,0,0,-)\rangle\,, \\
  |11\rangle & |\mathrm{D}_1^1 \,[222]\,(-,-,+,0,0,+)\rangle\,, & 
    |12\rangle & |\mathrm{C}_1   \,[222]\,(+,+,0,-,0,-)\rangle\,, \\
  |13\rangle & |\mathrm{C}_1   \,[222]\,(+,0,+,0,-,-)\rangle\,, & 
    |14\rangle & |\mathrm{C}_1   \,[222]\,(+,-,+,0,0,-)\rangle\,, \\
  |15\rangle & |\mathrm{C}_1   \,[222]\,(+,+,0,0,-,-)\rangle\,, & 
    |16\rangle & |\mathrm{D}_3^0 \,[303]\,(+,-,+,-,+,-)\rangle\,, \\
  |17\rangle & |\mathrm{D}_1^0 \,[303]\,(+,+,-,-,-,+)\rangle\,, & 
    |18\rangle & |\mathrm{C}_1   \,[303]\,(+,+,-,+,-,-)\rangle\,. 
 \end{array}   
$$
Each orbit gives rise to only one state $|j; A_1\rangle\equiv|j'\rangle$,
$j=1,2,\dots,18$, with symmetry $\Gamma=A_1$ constructed as a normalized
sum of all $|\mathrm{D}_6|/|U|$ states in a given orbit 
$O(U,[k],\mu)\equiv O_j$.

\setlength{\unitlength}{0.23mm}

Since the Hamiltonian considered is isotropic, then we can use the total 
spin number $S$ to label eigenstates. The operator $\mathbf{S}^2$ can 
be expressed as
 $$
 \mathbf{S}^2=\sum_{i=1}^6 \mathbf{s}_i^2+2\sum_{i=1}^5\sum_{i'=i+1}^6
   \mathbf{s}_i\cdot \mathbf{s}_{i'}=
 12+2\mathcal{H}+2\sum_{i=1}^6 \mathbf{s}_i\cdot \mathbf{s}_{i+2}+
   2\sum_{i=1}^3 \mathbf{s}_i\cdot \mathbf{s}_{i+3}\,.
 $$
 The diagonal parts of $\mathcal{H}$ and $\mathbf{S}^2$ are given as
 \begin{eqnarray*}
   \mathcal{H}_{\mathrm{diag}} &=& \sum_{i=1}^6 s^z_i s^z_{i+1}\,, \\
 \mathbf{S}^2_{\mathrm{diag}} &=& 12+2\mathcal{H}_{\mathrm{diag}}
   +2\sum_{i=1}^6 s^z_i s^z_{i+2} +2\sum_{i=1}^3 s^z_i s^z_{i+3}\,.
 \end{eqnarray*}
 In the first case (for $\mathcal{H}_{\mathrm{diag}}$) one obtains
the following 18 eigenvalues: 0, 0, --1, 0, --2, --2, --2, --2, 2, --1, --1, 
0, 0, --3, 1, --6, 2, and --2. The case of $\mathbf{S}^2$ operator is even 
more simple, since the diagonal values depend only on the partition $[k]$
and one obtains four numbers: 12, 10 (orbits 2, 3, 4), 
8 (orbits 5 through 15), and 6 (orbits 16, 17, 18).

To consider off-diagonal terms containing step operators $s_i^\pm$ we
start with a graph of partitions:
  \begin{center}
  \begin{picture}(245,45)
   \put( 10,20){\circle*{5}}
   \put( 85,20){\circle*{5}}
   \put(160,20){\circle*{5}}
   \put(235,20){\circle*{5}}
   \put( 10, 0){\makebox(0,0){[060]}}
   \put( 85, 0){\makebox(0,0){[141]}}
   \put(160, 0){\makebox(0,0){[222]}}
   \put(235, 0){\makebox(0,0){[303]}}
   \put( 20,20){\line(1,0){55}}
   \put( 95,20){\line(1,0){55}}
   \put(170,20){\line(1,0){55}}
   \put( 85,35){\circle{30}}
   \put(160,35){\circle{30}}
  \end{picture}
  \end{center}
This graph leads a block structure of off-diagonal parts of the Hamiltonian 
and $\mathbf{S}^2$ (symmetric) matrices:
 $$
  \left(\begin{array}{*{4}{c}}
   \mathbf{0}  & \mathcal{C} & \mathbf{0}  & \mathbf{0}  \\
   \mathcal{C}^\dag & \mathcal{A} & \mathcal{D} & \mathbf{0}  \\
   \mathbf{0}  & \mathcal{D}^\dag & \mathcal{B} & \mathcal{E} \\
   \mathbf{0}  & \mathbf{0}  & \mathcal{E}^\dag & \mathbf{0}  
  \end{array}\right), 
 $$
 where rows and columns are labeled by non-ordered partitions and $\dag$ 
denotes the Hermitian conjugation. All these blocks may contain non-zero 
elements for $\Gamma=A_1$, because this representation appears in the 
decomposition of any transitive irreps. More detailed structure can be 
obtained if orbit stabilizers are taken into account (now rows and columns 
are labeled by pairs $(U,[k])$; borders of the blocks introduced above are 
denoted by lines):
 $$
  \left(\begin{array}{c|cc|cccc|ccc}
   \mathbf{0}  & \mathcal{C}_1 & \mathcal{C}_2 
      & \multicolumn{4}{c|}{\mathbf{0}} & & \mathbf{0} &  \\
  \hline
   \mathcal{C}^\dag_1 & \mathcal{A}_1 & \mathcal{A}_3 
    & \mathcal{D}_1 & \mathcal{D}_2 & \mathcal{D}_3 & \mathcal{D}_4 
    & & & \\[-7pt]
    & & & & & & & & \mathbf{0} & \\[-7pt]
   \mathcal{C}^\dag_2 & \mathcal{A}^\dag_3 & \mathcal{A}_2 
    & \mathcal{D}_5 & \mathcal{D}_6 & \mathcal{D}_7 & \mathcal{D}_8 
    & & & \\
  \hline
   & \mathcal{D}^\dag_1 & \mathcal{D}^\dag_5 
   & \mathcal{B}_1 & \mathcal{B}_5 & \mathcal{B}_6 & \mathcal{B}_7 
   & \mathcal{E}_1 & \mathcal{E}_2 & \mathcal{E}_3 \\
   & \mathcal{D}^\dag_2 & \mathcal{D}^\dag_6 
   & \mathcal{B}^\dag_5 & \mathcal{B}_2 & \mathcal{B}_8 & \mathcal{B}_9 
   & \mathcal{E}_4 & \mathcal{E}_5 & \mathcal{E}_6 \\[-7pt]
 \mathbf{0}  & &&&&&&&& \\[-7pt]
   & \mathcal{D}^\dag_3 & \mathcal{D}^\dag_7 
   & \mathcal{B}^\dag_6 & \mathcal{B}^\dag_8 & \mathcal{B}_3 
               & \mathcal{B}_{10} 
   & \mathcal{E}_7 & \mathcal{E}_8 & \mathcal{E}_9 \\
   & \mathcal{D}^\dag_4 & \mathcal{D}^\dag_8 
   & \mathcal{B}^\dag_7 & \mathcal{B}^\dag_9 & \mathcal{B}^\dag_{10} 
              & \mathcal{B}_4 
   & \mathcal{E}_{10} & \mathcal{E}_{11} & \mathcal{E}_{12} \\
  \hline
   & & & \mathcal{E}^\dag_1 & \mathcal{E}^\dag_4 & \mathcal{E}^\dag_7 
          & \mathcal{E}^\dag_{10} & \\
\mathbf{0}  & \multicolumn{2}{c|}{\mathbf{0}}  
  & \mathcal{E}^\dag_2 & \mathcal{E}^\dag_5 & \mathcal{E}^\dag_8 
        & \mathcal{E}^\dag_{11} 
    & & \mathbf{0}  \\
   & & & \mathcal{E}^\dag_3 & \mathcal{E}^\dag_6 & \mathcal{E}^\dag_9 
           & \mathcal{E}^\dag_{12} &   
  \end{array}\right), 
 $$
 Depending on the representation considered some blocks have to contain
only zeroes. For example, when $\Gamma=A_2$, then only blocks  
$\mathcal{A}_2$, $\mathcal{B}_1$, $\mathcal{B}_4$, $\mathcal{B}_7$,  
$\mathcal{D}_5$, $\mathcal{D}_8$, $\mathcal{E}_3$, $\mathcal{E}_{12}$ 
may contain non-zero entries. It follows from the fact that the irrep
$A_2$ appears in the decomposition of a transitive representation 
$R^{\mathrm{D}_N:U}$ if $U$ is a cyclic subgroup of $\mathrm{D}_N$.

The classification of orbits with $M=0$  by type $\widetilde{U}$ 
yields that there are six different types of orbits, namely
$\mathrm{D}_6, \mathrm{D}_3^0, \mathrm{C}_2, \mathrm{D}_1^0, \mathrm{D}_1^1, 
\mathrm{C}_1$, so there are 21 different double cosets to be considered.. 
However, the matrix structure presented above excludes cosets
$\mathrm{D}_6 g\mathrm{D}_6$,
$\mathrm{D}_6 g\mathrm{D}_3^0$,
$\mathrm{D}_6 g\mathrm{D}_1^1$,
$\mathrm{D}_6 g\mathrm{C}_2$, and
$\mathrm{D}_3^0 g\mathrm{D}_3^0$. 
 
Two notes are in place: (i) since we consider the case $\Gamma=A_1$, so
all coefficients $B^{UV\Gamma}=1$; (ii) the off-diagonal terms contains
products $s_i^+s_{i'}^-$, which acting on kets $|m_im_{i'}\rangle$ gives
zero or $2|m_i+1\:m_{i'}-1\rangle$, therefore terms 
$(s_i^+s_{i'}^-+s_i^-s_{i'}^+)/2$ are replaced by 
$\sigma_i^+\sigma_{i'}^-+\sigma_i^-\sigma_{i'}^+$, where 
$\sigma_i^{\pm1}|\pm1\rangle=|\rangle$ and 
$\sigma_i^\pm|m_i\rangle=|m_i\pm1\rangle$, otherwise.

Let us consider, as an example, the 16th column of an operator matrix.
This column corresponds to the two-element orbit containing the N\'eel
configurations $|+,-,+,-,+,-\rangle$ and $|-,+,-,+,-,+\rangle$
with a stabilizer $\mathrm{D}_3^0$ and labeled by the non-ordered
partition $[303]$. Therefore, non-zero entries in this column, for 
operators constructed of bilinear terms, can appear only in the block 
$\mathcal{E}$, {\it i.e.}\ rows labeled by the partition $[222]$ (except for 
the diagonal terms $\sum_{i,i'} s_i^zs_{i'}^z$). So, one has to consider four 
sets of double cosets $\mathrm{D}_3^0 \backslash\mathrm{D}_6/U$ with 
$U=\mathrm{C}_2$, $\mathrm{D}_1^0$, $\mathrm{D}_1^1$, and $\mathrm{C}_1$.
In all cases $k=3$ and $\kappa=3$; in the first case ($U=\mathrm{C}_2$)
$l=2$, $\lambda=3$, so $\mathrm{sd}(k,l)=6$ and $\chi=1$, whereas in
the other cases we have $l=1$, $\lambda=6$, $\mathrm{sd}(k,l)=3$ and
$\chi=2$. Eqs.\ (\ref{Hgdcd}), (\ref{h1}), and (\ref{h4}) 
yield the following formulas (the repetition indices $r$, $s$ may be 
omitted):
 \begin{eqnarray*}
 h_{\mathrm{C}_2 \mu,\mathrm{D}_3^0 \nu}(A_1)\; =\;2\sqrt3\: f(E) \,; &&
 h_{\mathrm{C}_1 \mu ,\mathrm{D}_3^0 \nu }(A_1) 
    \;=\; \sqrt6\: \Big(f(E)+f(C_6)\Big) \,;  \\
 h_{\mathrm{D}_1^1\mu,\mathrm{D}_3^0\nu}(A_1) \;=\; 2\,\sqrt3\: f(E)\,; &&
 h_{\mathrm{D}_1^0\mu,\mathrm{D}_3^0\nu}(A_1) \;=\;
       \sqrt3\:\Big(f(E) + f(C_6)\Big)\,. 
 \end{eqnarray*}
  Therefore, one has to take into account the states $|j\rangle$, for 
$5\le j\le15$ and $P(C_6)|j\rangle$ for $j=6,7,8,12,13,14$, and 15. Only two 
of them are obtained from $|16\rangle$ under the action of bilinear terms 
$\sigma_i^+\sigma_{i'}^-$, namely 
 $$
   P(C_6)|7\rangle=|+,0,+,-,0,-\rangle=\sigma_2^+\sigma_5^-
  |+,-,+,-,+,-\rangle\,, 
$$
 and 
 $$
   |14\rangle=|+,-,+,0,0,-\rangle=\sigma_4^+\sigma_5^-
  |+,-,+,-,+,-\rangle\,. 
$$
  Hence, the non-zero off-diagonal terms in the sixteenth columns are
$$
  \mathcal{H}_{14,16}=\sqrt6\,;\qquad
  \mathbf{S}^2_{7,16}=2\sqrt3\,,\quad
  \mathbf{S}^2_{14,16}=2\sqrt6\,.\quad
$$
 All other elements of operator matrices can be obtained using the
following formulas:
 \begin{eqnarray*}
 h_{\mathrm{C}_1 \mu ,\mathrm{D}_6 \nu }(A_1) = \sqrt2\; 
 h_{\mathrm{D}_1^0 \mu,\mathrm{D}_6 \nu}(A_1) & =&2\sqrt3\; f(E) \,;  \\
 h_{\mathrm{D}_1^0\mu,\mathrm{C}_2\nu}(A_1) = 
 h_{\mathrm{D}_1^1\mu,\mathrm{C}_2\nu}(A_1) =
 h_{\mathrm{D}_1^1\mu,\mathrm{D}_1^0\nu}(A_1) 
  &=&2\,\sum_{d=0}^2 f(C_6^d)\,, \\
 h_{\mathrm{C}_1\mu,\mathrm{C}_2\nu}(A_1) = \sqrt2\;
 h_{\mathrm{C}_2\mu,\mathrm{C}_2\nu}(A_1) &=& \sqrt2\;\sum_{d=0}^2\Big(
  f(C_6^d)+f(U_d)\Big)\,; \\
 h_{\mathrm{D}_1^0\mu,\mathrm{D}_1^0\nu}(A_1)=
 h_{\mathrm{D}_1^1\mu,\mathrm{D}_1^1\nu}(A_1)&=&f(E)+f(C_6^3)+
2\Big(f(C_6)+f(C_6^2)\Big)\,;\\
 h_{\mathrm{C}_1\mu,\mathrm{D}_1^0\nu}(A_1) =
 h_{\mathrm{C}_1\mu,\mathrm{D}_1^1\nu}(A_1) 
  &=&\sqrt2\,\sum_{d=0}^5 f(C_6^d)\,; \\
 h_{\mathrm{C}_1\mu,\mathrm{C}_1\nu}(A_1) 
  &=&\sum_{d=0}^5 \Big(f(C_6^d)+f(U_d)\Big)\,. \\
 \end{eqnarray*}
  The matrices of the Hamiltonian $\mathcal{H}$ and the square
of total spin $\mathbf{S}^2$ are presented in Table~\ref{hs2}.

\begin{table}
\caption{The Hamiltonian (the upper triangle) and $\mathbf{S}^2/2$
(the lower triangle) matrices for $N=6$ spins $s=1$ in the basis 
$\{|j'\rangle\equiv|j;A_1\rangle\mid 1\le j\le 18\}$\label{hs2} 
(the lines are borders of blocks $\mathcal{A}$,
$\mathcal{B}$, $\mathcal{C}$, $\mathcal{D}$, and $\mathcal{E}$)}
$$
\advance\arraycolsep -1.5pt
 \begin{array}{r|rrr|*{10}{r}r|rrr}
   \cdot & \cdot & 2\sqrt3 & \cdot & \cdot & \cdot & \cdot & \cdot 
     & \cdot & \cdot & \cdot & \cdot & \cdot & \cdot & \cdot & \cdot 
    & \cdot & \cdot \\  \hline
\multicolumn{1}{c|}{} & \cdot & \cdot & 2\sqrt2 & \cdot & \cdot 
  & \cdot & \cdot & \cdot 
 & \cdot & \cdot & \cdot & \cdot & \sqrt2 & \sqrt2 & \cdot & \cdot 
  & \cdot\\ \cline{2-2}
    \multicolumn{2}{c|}{} & -1 & 2 & \sqrt2 & \cdot & \sqrt2 & \cdot 
  & \cdot & \sqrt2 & \sqrt2
  & \cdot & \cdot & 2&\cdot&\cdot&\cdot&\cdot \\ \cline{1-1}\cline{3-3}
6 & \multicolumn{2}{c|}{}  & \cdot & \cdot & \sqrt2 & \cdot & \sqrt2 & \cdot 
   & \cdot & \cdot & 1 & 1 & \cdot & \cdot & \cdot & \cdot & \cdot \\
    \cline{1-2}  \cline{4-18}
\sqrt6 & 5 & \multicolumn{2}{|c|}{}  
 & -2 & 2 & \cdot & 2 & \cdot & \cdot & \cdot & \cdot & \cdot & \cdot 
    & \cdot & \cdot & \cdot & \cdot \\ \cline{3-3}\cline{5-5}
2\sqrt3 & 2\sqrt2 & 7 & \multicolumn{2}{|c|}{}  
 & -2 & \cdot & \cdot & \cdot & \cdot & \cdot & \cdot & \cdot & \sqrt2 
     & \cdot & \cdot & \cdot & \cdot \\ \cline{4-4} \cline{6-6}
2\sqrt3 & 2\sqrt2 & 4 & 7 & \multicolumn{2}{|c|}{}   &
 -2 & \cdot & \cdot & \cdot & \cdot & \sqrt2 & \sqrt2 & \cdot & \cdot 
  & \cdot & \cdot & \cdot \\ \cline{1-5} \cline{7-7}
\cdot & \cdot & \sqrt2 & \sqrt2 & 4 & \multicolumn{2}{|c|}{} &
 -2 & \cdot & \cdot & \cdot & \cdot & \cdot & \sqrt2 & \cdot & \cdot 
  & \cdot & \cdot \\ \cline{6-6}\cline{8-8}
\cdot & \cdot & \sqrt2 & \sqrt2 & 2 & 4 & \multicolumn{2}{|c|}{} &
 2 & \cdot & \cdot & \sqrt2 & \sqrt2 & \cdot & \cdot & \cdot & \cdot 
  & \cdot \\ \cline{7-7}\cline{9-9}
\cdot & 2 & \sqrt2 & \cdot & \cdot & \cdot & 4 & \multicolumn{2}{|c|}{} &
  -1 & \cdot & \sqrt2 & \cdot & \cdot & \cdot 
 & \cdot & \cdot & \sqrt2 \\ \cline{8-8}\cline{10-10}
\cdot & \cdot & \sqrt2 & \sqrt2 & 2 & \cdot & \cdot & 4 
 & \multicolumn{2}{|c|}{} & -1 & \cdot & \sqrt2 & \cdot & \cdot & \cdot 
 & \cdot & \sqrt2 \\ \cline{9-9}\cline{11-11}
\cdot & 2 & \cdot & \sqrt2 & \cdot & \cdot & \cdot & \cdot & 4
 & \multicolumn{2}{|c|}{} & \cdot & \cdot & \cdot & 1 & \cdot & \cdot &
  \cdot \\ \cline{10-10}\cline{12-12}
\cdot & \cdot & \sqrt2 & \sqrt2 & 2 & \cdot & \cdot & 2 & \cdot & 4 
 & \multicolumn{2}{|c|}{} & \cdot & \cdot & 1 & \cdot & \cdot & \cdot \\
 \cline{11-11}\cline{13-13} 
\cdot & \cdot & \sqrt2 & \sqrt2 & 2 & 2 & \cdot & \cdot & \cdot & \cdot & 4
 & \multicolumn{2}{|c|}{} & -3 & \cdot & \sqrt6 & \cdot & 1 \\\cline{12-12}
 \cline{14-14}
\cdot & \sqrt2 & 1 & 2 & \cdot & \sqrt2 & \sqrt2 & \cdot & \sqrt2 & \sqrt2
  & \cdot & 5 & \multicolumn{2}{|c|}{} & 1 & \cdot & \sqrt2 & 1 \\
 \cline{13-13}\cline{15-18}
\cdot & \sqrt2 & 1 & 2 & \cdot & \cdot & \sqrt2 & \sqrt2 & \sqrt2 & \cdot
 & \sqrt2 & \cdot & 5 & \multicolumn{2}{|c|}{} &  -6 & \cdot & \cdot
\\ \cline{14-14}\cline{16-16}
\cdot & \sqrt2 & 3 & \cdot & \cdot & \sqrt2 & 2\sqrt2 & \sqrt2 & \cdot 
 & \cdot & \cdot & 1 & 1 & 6 & \multicolumn{2}{|c|}{} & 2 & \cdot \\
 \cline{15-15} \cline{17-17}
\cdot & \sqrt2 & 1 & 2 & \cdot & \cdot & \cdot & \cdot & 2\sqrt2 & \sqrt2
 & \sqrt2 & 2 & 2 & \cdot & 4 & \multicolumn{2}{|c|}{} & -2 \\ \cline{1-16}
 \cline{18-18}
\cdot & \cdot & \cdot & \cdot & \cdot & \cdot & \sqrt3 & \cdot & \cdot 
  & \cdot & \cdot & \cdot & \cdot & \sqrt6 & \cdot & 3 & 
  \multicolumn{2}{|c}{}  \\ \cline{17-17}
\cdot & \cdot & \cdot & \cdot & \cdot & \cdot & 1 & \cdot & 2 & \cdot 
  & \cdot & \sqrt2 & \sqrt2  & \cdot & \sqrt2 & \cdot & 3 & 
  \multicolumn{1}{|c}{} \\ \cline{18-18}
\cdot & \cdot & \cdot & \cdot & \sqrt2 & \sqrt2 & \cdot & \sqrt2 & \cdot
 & \sqrt2 & \sqrt2 & 1 & 1 & 1 & 1 & \cdot & \cdot & 3 
 \end{array}
$$
\end{table}

The irrep $A_1$ has not been chosen merely due to the simplest form
of the coefficients $B^{UVA_1}(g)$, but in order to illustrate the 
other way of determination of matrix elements in this case. Let an
orbit $O_j$ consists of $n_j=|G|/|U_j|$ elements ($U_j$ is a stabilizer
of the configuration $|j\rangle\equiv|U_j\,[k]\,\mu\rangle$). Hence,
the fully symmetric state $|j'\rangle\equiv|j;\Gamma_0\rangle$ is
given as
 \begin{equation}\label{gam0sta}
   |j'\rangle=\frac{1}{\sqrt{n_j}}\,\sum_{i=1}^{n_j} P(g_i)|j\rangle\,,
 \end{equation}
 where $g_i$ are representatives of left cosets $g_iU_j\in G/U_j$.
Since a stabilizer of each element in the orbit $O_j$ has $|U_j|$ elements
than  
 $$
  |j'\rangle = \frac{1}{|U_j|\sqrt{n_j}} \sum_{g\in G} P(g_i)|j\rangle
  = \frac{\sqrt{n_j}}{|G|}\sum_{g\in G} P(g)|j\rangle \,.
 $$
 Let $\mathcal{H}'$ denote the off-diagonal part of a Hamiltonian
$\mathcal{H}$, {\it i.e.}\ $\mathcal{H}'=
\mathcal{H}-\mathcal{H}_{{\rm diag}}$.
The symmetry operators $P(g)$ commute with a Hamiltonian $\mathcal{H}'$,
so 
  $$
   \mathcal{H}'|j'\rangle 
 = \frac{\sqrt{n_j}}{|G|}\:\sum_{g\in G} P(g)\mathcal{H}'|j\rangle\,.
  $$
  To simplify the further considerations we limit ourselves to the cases
$s=1/2,1$, when $s_i^+s_{i'}^-|m_i\dots, m_{i'}\rangle=
|m_i+1,m_{i'}-1\rangle$ or $\sigma_i^+\sigma_{i'}^-|m_i,m_{i'}\dots\rangle
=|m_i+1, m_{i'}-1\rangle$, since in the case of larger spin numbers 
more detailed discussion is necessary. Under this assumption, each of  
off-diagonal terms of the bilinear Hamiltonian gives the empty sate 
$|\rangle$ or one configuration without any coefficients. Let $n_{jk}$
denote a number of configurations obtained from $|j\rangle$ (by the
action of $\mathcal{H}'$) and belonging to the orbit $O_k$. Then 
$\sum_g P(g)\mathcal{H}'|j\rangle$ contains $|G|n_{jk}$ configurations 
from the orbit $O_k$. Since the state $|k'\rangle\equiv|k;A_1\rangle$ are 
constructed from {\em all}\/ $n_k$ elements of the orbit $O_k$, then 
$|G|n_{jk}$ has to be divisible by $n_k$ and each element the orbit $O_k$ 
appears
$|G|n_{jk}/n_k$ times in the considered sum $\sum_gP(g)\mathcal{H}'
|j\rangle$. Therefore, one can construct $|G|n_{jk}/\sqrt{n_k}$ vectors
$|k'\rangle$. As the result we obtain
 $$
  \langle k'| \mathcal{H}'|j'\rangle = \frac{\sqrt{n_j}}{|G|}\:
  \frac{|G|n_{jk}}{\sqrt{n_k}} = n_{jk}\sqrt{\frac{n_j}{n_k}}=
   n_{jk}\sqrt{\frac{|U_k|}{|U_j|}}\,.
$$
 In the example presented above $|j\rangle=|+,-,+,-,+,-\rangle$ and the 
off-diagonal terms of the Hamiltonian give us six configurations belonging
to the orbit $O_{14}$ (with $n_{14}=12$). The additional terms in 
$\mathbf{S}^2$ operator lead to three configurations in the 6-element
orbit $O_7$. Therefore,
 $$
   \mathcal{H}'_{14,16}={1\over2}\mathbf{S}^2_{14,16}=6\sqrt{\frac{2}{12}}
 =\sqrt6\,;\qquad 
   {1\over2}\mathbf{S}^2_{7,16}=3\sqrt{\frac{2}{6}}=\sqrt3\,,
 $$
 what agrees with the previous results. 

The further considerations go beyond the scope of this paper,
since there are some numerical rather than combinatorial problems.
However, we are very close to the solution of the eigenproblem and it
is interesting to calculate the ground state energy. 
In the case of isotropic Hamiltonians, when $S$ is a good quantum number,
we determine an orthonormal basis for each space $L_{M=S,A_1}$ solving the 
eigenproblem
$\mathbf{S}^2|\phi\rangle = S(S+1)|\phi\rangle$ for $S=0,1,2,3,4,6$ 
($S=5$ may be omitted since there are no states with $M=S=5$ and 
$\Gamma=A_1$). It can be done, for example, using such packages as
{\it MATHEMATICA}$^{\mbox{\scriptsize\textregistered}}$ or numerical 
libraries for calculations with an arbitrary precision as {\sf GMP} 
\cite{gmp}. Then, we transform the Hamiltonian to this new basis, what leads 
to a quasi-diagonal matrix consisting of blocks with dimensions 5, 1, 6, 
2, 3, 1 for $S=0,1,2,3,4,6$, respectively. It is easy to solve the 
eigenproblems for such small blocks and, for example, in the case $S=0$ we 
obtain five eigenvalues: --8.6174, --4.7979, --3.3302, --0.3391, 3.0846, 
where the first one corresponds to the ground state with energy per spin 
--1.43624. In the case of such small eigenproblems one can also determine 
eigenstates\footnote{Due to its form the eigenproblem for $\mathbf{S}^2$ can 
be solved exactly in many cases. This leads to an {\em exact}\/ form of 
blocks in the Hamiltonian matrix. However, the Hamiltonian eigenproblem,
except for some special cases, has to be solved numerically with the precision
depending on software and hardware used. Therefore, all numbers obtained
give only approximate values of energies, correlations, vector 
coefficients {\it etc}. } and, therefore, it is possible to calculate spin 
correlations in the ground state. 

\section{Final remarks}
 In this paper we have studied properties of the dihedral groups 
$\mathrm{D}_N$ to obtain analytical formulas for number of double cosets and 
their elements. These results are necessary to exploit in full symmetry 
properties of a given spin operator and to obtain its matrix elements in 
the symmetry adapted basis. The groups considered are important in 
investigation of magnetic macromolecules forming rings or other structures 
with isomorphic symmetry group ({\it e.g.}\ crown-like structures). In a 
sense this paper, together with the preceding one \cite{prev}, continues 
works done for the cyclic groups by T.~Lulek and one of the authors (WF) 
\cite{cyclic}. In those articles the arithmetic structure of a number of 
nodes $N$ was mostly taken into account and some combinatorial object were
considered. The present series of papers is mainly devoted to combinatorial 
and group-theoretical objects and the most important result is the formula 
for matrix elements expressed as the sum over representatives of double 
cosets. In the case of dihedral groups the approach proposed has two 
advantages: (i) all formulas can be obtained in analytical form and they
can be used in computer programmes; (ii) in one dimension a loop over double 
coset representatives is, in general, shorter than a loop over all 
off-diagonal elements of bilinear Heisenberg Hamiltonian, what has been 
discussed in the previous paper \cite{prev}; moreover, it is easy to modify 
Hamiltonian introducing, for example, interactions with the next nearest 
neighbors or alternating exchange integrals. 

It should be underlined, however, that the final diagonalization of a given 
Hamiltonian is (almost) always done numerically, with all round-off errors 
caused by the float-point representation of numbers. Secondly, the method 
proposed only slightly increase the size of system which can be investigated. 
The so-called `combinatorial explosion' leads to very large eigenproblems 
for quite small systems: {\it e.g.}\ a system consisting of twelve spins 
$s=2$ yields more than 240 millions Ising configurations, so there are about 
ten millions orbits of the dihedral group $\mathrm{D}_{12}$; more than 800 
thousands orbits correspond to $M=0$, so this is the approximate of 
$\mathbf{S}^2$ eigenproblem dimension.  On the other hand, not so much smaller 
system with $N=8$ leads to about 2,500 orbits with $M=0$, what can be 
investigated using modern computers. A number in the same range is obtained
for $N=19$ and $s=1/2$, but we hope to determine the ground
state energy for the Heisenberg antiferromagnetic ring with $N=24$ nodes, 
where a system of about 50 thousands linear equations has to be solved.
However, if one would like to determine the thermodynamic properties then
{\em all}\/ states have to be taken account, among others these with 
the symmetry $\Gamma=E_p$, which are double-fold degenerated, so the 
dimension of $\mathbf{S}^2$ eigenproblem is approximately two times larger 
than that for one-dimensional irreps.

\section*{Acknowledgments}
This work is partially supported by the State Committee for Scientific 
Research (KBN) within the project No 2~P03B~074~19. The authors are indebted
to Prof.~G.~Kamieniarz for fruitful discussions. 

\clearpage

\section*{Figure captions}
\begin{description}
\item{Figure 1:~~~}
  Numbering of ring nodes and elements of the group 
    $\mathrm{D}_6$ (the six-fold axis $C_6$ is denoted by the full 
    hexagon)
  \item{Figure 2:~~~} Lattice of conjugated subgroups for the dihedral group 
   $\mathrm{D}_6$; lines correspond to the group-subgroup relations
\item{Figure 3:~~~} Orbit representatives for $N=6$, $s=1$ and $M=0$; full 
and empty circles denote $m_j=+1,-1$, respectively
\end{description}

\end{document}